# De la recherche sociale d'information à la recherche collaborative d'information

Victor Odumuyiwa

Equipe SITE-LORIA, Nancy Université
Laboratoire Lorrain de Recherche en Informatique et ses Applications
Campus Scientifique – BP 239 54506 Vandœuvre lès Nancy Cedex, France
victor.odumuyiwa@loria.fr

**RESUME**
Dans ce papier, nous expliquons la recherche sociale d'information (RSI) et la recherche collaborative d'information (RCI). Nous considérons la RSI comme un processus de recherche d'un collaborateur potentiel pour la résolution d'un problème informationnel. En revanche la RCI implique la compréhension et la résolution mutuelle d'un problème informationnel par les collaborateurs. Nous nous sommes intéressés au passage de la RSI à la RCI donc nous avons développé un modèle de communication pour faciliter le partage de connaissance pendant la RCI.
**MOTS-CLES** : Recherche collaborative d'information, recherche sociale d'information, partage de connaissance, communication interpersonnelle, intelligence collective.

**ABSTRACT**
In this paper, we explain social information retrieval (SIR) and collaborative information retrieval (CIR). We see SIR as a way of knowing who to collaborate with in resolving an information problem while CIR entails the process of mutual understanding and solving of an information problem among collaborators. We are interested in the transition from SIR to CIR hence we developed a communication model to facilitate knowledge sharing during CIR.

La RSI est fondée sur le fait que l'on ne peut pas séparer le producteur d'une information de son produit (Kirsch, 2005) et que nous utilisons les gens pour trouver des contenus et les contenus pour trouver des gens (Morville, 2002). Dans le contexte de notre travail, la RSI est composé de trois éléments : utilisateur (U), problème informationnel (P) et activité (A) de recherche d'information.

$RSI = \{U, P, A\}$

Les compétences (savoir et savoir faire) mises en œuvre lors des activités dans la RSI par les utilisateurs constituent l'intelligence collective de la communauté. Les activités ne sont pas forcément centrées sur un problème partagé par l'ensemble de la communauté. Il s'agit donc pour nous ce que nous considérons comme des activités de coopération (Odumuyiwa & David 2009). En revanche la mémoire de ces activités permet de trouver des collaborateurs potentiels pour la résolution d'un nouveau problème similaire à un autre déjà résolu par la communauté.

Nous nous sommes intéressés au passage de la RSI à la RCI. Ce passage commence par le partage du problème informationnel P' entre l'utilisateur l'ayant formulé et les collaborateurs potentiels identifiés dans la communauté. Dans le cadre de la RCI, U' représente l'ensemble des utilisateurs qui collaborent pour résoudre le problème informationnel partagé P'. Leurs activités A' sont mutuellement effectuées et centrées autour de P'.

RCI = {U', P', A'}

Un problème partagé n'implique pas forcément une compréhension partagée du problème. C'est pour cette raison que le point de départ dans le passage de la RSI à la RCI est la définition et la clarification collaborative du problème informationnel. Pour arriver à une compréhension partagée du problème, les utilisateurs collaborant entament un processus d'intégration et de différentiation de leur compréhension du problème à travers des annotations et de la communication interpersonnel.

Nous avons développé un modèle de communication pour gérer ce processus. Ce modèle consiste en quatre composants qui définissent un contexte collaboratif : contexte, objet, expéditeur, et destinataire. Chaque échange dans la collaboration se fait dans un contexte particulier. Pour nous, le *contexte* intègre le problème informationnel, l'objectif de recherche et la période de la définition du problème. L'*objet* de la communication est la connaissance exprimée par un acteur qui a comme attributs : le contenu, la structure du contenu, le format, et la direction qui peut être unidirectionnel quand il s'agit d'un objet envoyé par un expéditeur à un ou plusieurs destinataires sans attente de retour ou conversationnel quand il s'agit d'un objet d'un expéditeur à un ou plusieurs destinataires avec un retour suite à l'expédition.

L'*expéditeur*, qui peut être aussi un *destinataire*, a comme attributs : son profil qui consiste en son identifiant, nom, prénom, ses préférences ; sa localisation puisqu'il s'agit de collaboration à distance, ses réseaux, ses compétences et ses connaissances du domaine qui sont dérivées de ses activités.

Ce modèle permet d'acquérir les compétences employées lors d'une collaboration. Il sert également à gérer l'interaction entre les utilisateurs et faciliter le passage de la RSI à la RCI. Toutes les connaissances sont stockées pour une réutilisation.

## Bibliographie